\documentclass[%
aip,
cp, 
amsmath,amssymb,
reprint,%
]{revtex4-2}

\def\aap{\ {A\&A}\ }

\def\apj{\ {ApJ}\ }

\def\apss{\ {Ap\&SS}\ }

\def\mnras{\ {MNRAS}\ }

\def\pasj{\ {Publ. Astr. Soc. Japan}\ }

\usepackage{graphicx}
\usepackage{dcolumn}
\usepackage{bm}

\usepackage[utf8]{inputenc}
\usepackage[T1]{fontenc}
\usepackage{mathptmx} 

\def\lteq{\ {\raise-.5ex\hbox{$\buildrel<\over-$}}\ }
\def\apgt{\ {\raise-.5ex\hbox{$\buildrel>\over\sim$}}\ }
\def\aplt{\ {\raise-.5ex\hbox{$\buildrel<\over\sim$}}\ }
\def\lt{\ {\raise-.5ex\hbox{$\buildrel>$}}\ }
\def\gt{\ {\raise-.5ex\hbox{$\buildrel<$}}\ }
\def\eqgt{\ {\raise-.5ex\hbox{$\buildrel>\over-$}}\ }
\def\eqlt{\ {\raise-.5ex\hbox{$\buildrel<\over-$}}\ }

\usepackage{amsmath} 
\newcommand{\angstrom}{\textup{\AA}}

\begin{document}

\title{The paradox of infinitesimal granularity: \\
       Chaos and the reversibility of time in
       Newton's theory of gravity}
\author{Simon Portegies Zwart}
\email[Corresponding author:]{spz@strw.leidenuniv.nl}
\affiliation{Leiden Observatory, Leiden University, PO Box 9513, 2300 RA, Leiden, The Netherlands}
\author{Tjarda Boekholt}
\email[]{tjarda.boekholt@physics.ox.ac.uk}

\affiliation{Rudolf Peierls Centre for Theoretical Physics, Clarendon Laboratory, Parks Road, Oxford, OX1 3PU, UK}

\date{\today} 

\begin{abstract}
The fundamental laws of physics are time-symmetric, but our
macroscopic experience contradicts this. The time reversibility
paradox is partly a consequence of the unpredictability of Newton's
equations of motion. We measure the dependence of the fraction of
irreversible, gravitational N-body systems on numerical precision and
find that it scales as a power law. The stochastic wave packet
reduction postulate then introduces fundamental uncertainties in the
Cartesian phase space coordinates that propagate through classical
three-body dynamics to macroscopic scales within the triple's
lifetime. The spontaneous collapse of the wave function then drives
the global chaotic behavior of the Universe through the superposition
of triple systems (and probably multi-body systems). The paradox of
infinitesimal granularity then arises from the superposition
principle, which states that any multi-body system is composed of an
ensemble of three-body problems.
\end{abstract}

\maketitle

During the Great Plague of London in 1665, Isaac Newton went in
quarantine for two years at his home in Woolsthorpe. There he laid
down the foundation for his theory on gravitation. During last year's
COVID-19 pandemic, we have been working on one of the consequences of
Newton's theory of gravitation and it's relation to chaos and time
reversibility.

\begin{figure}[ht]
\includegraphics[width=0.45\textwidth,angle=-0.0]{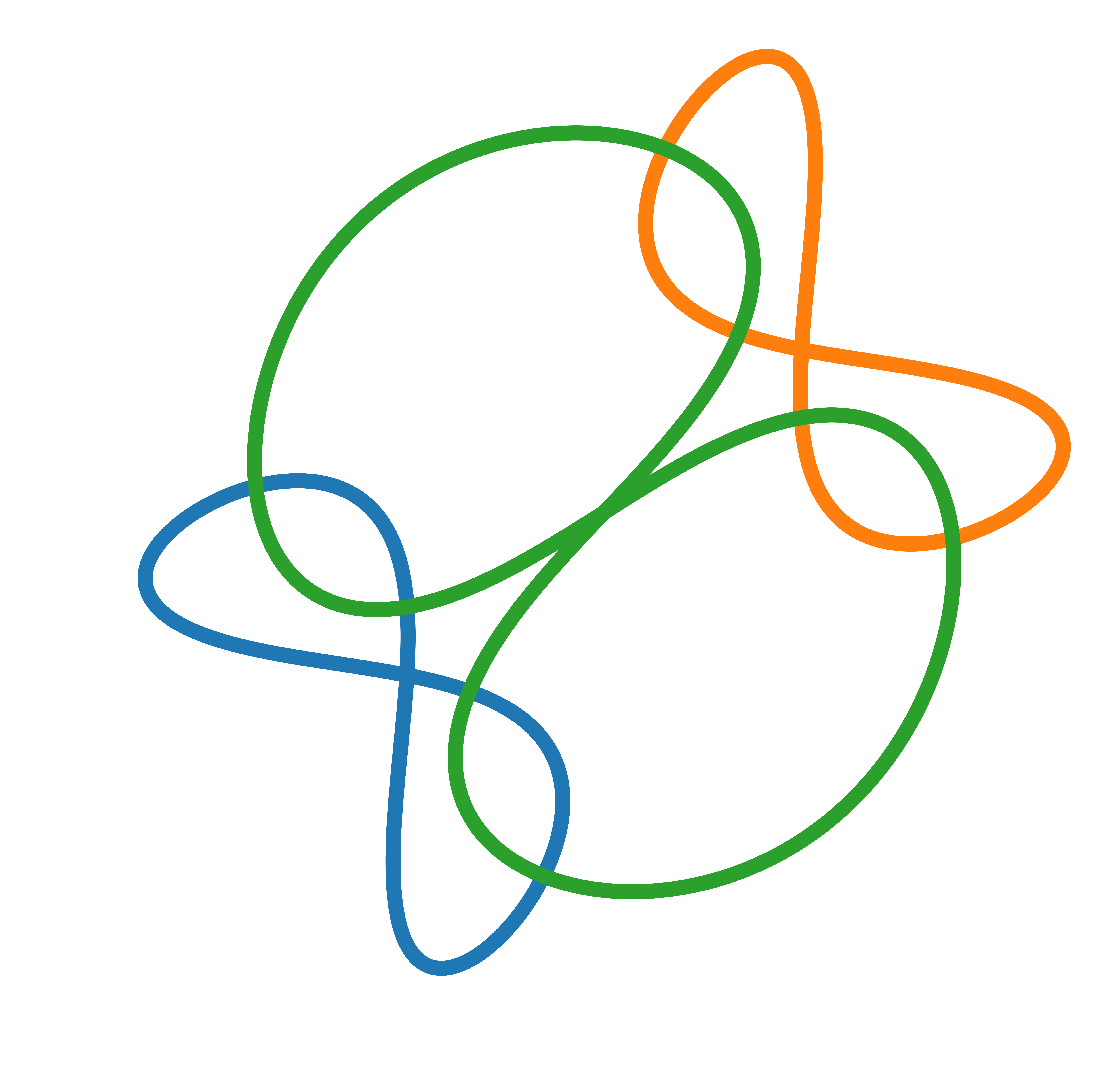}
~\includegraphics[width=0.45\textwidth,angle=-0.0]{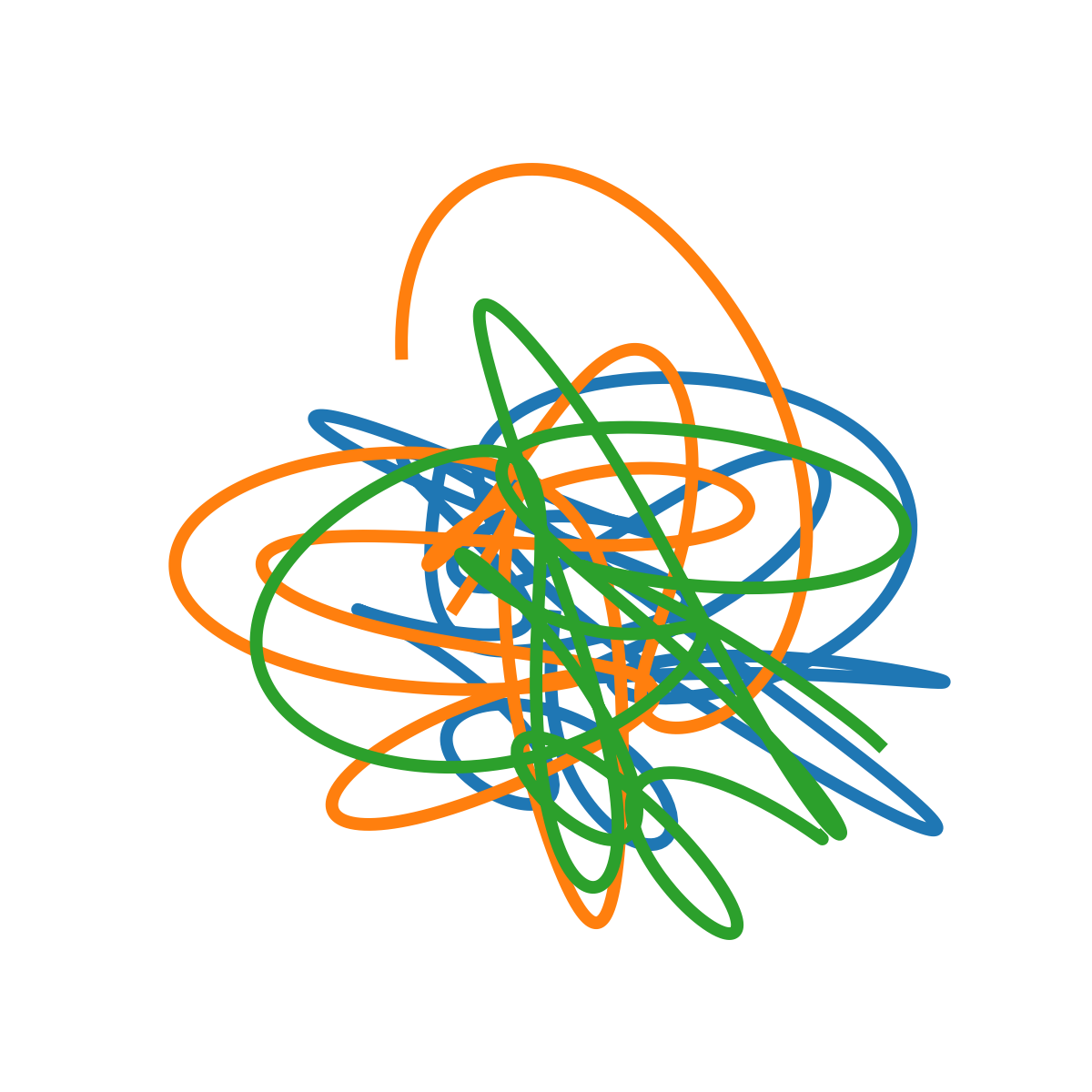}
\caption[]{Two orbits of three-body problems, a periodic braid to the
left, and right a partial solution of a democratic triple system.
The orbit to the left is bounded and time reversible, it was found
by \cite{2018PASJ...70...64L}. The orbit to the right is unbounded
and not time reversible. }
\label{fig:braids}
\end{figure}

The chaotic problem of three bodies started with Henry Poincar{\'e}
who argued in 1891 \cite{1891BuAsI...8...12P} that the problem was
unsolvable, but he was wrong. This led to the Kolmogorov-Arnold-Moser
theorem, and a solution based on an infinite time-series
\cite{zbMATH02627145,zbMATH02639473}, which therefore has little
practical use \cite{PHSC_2001__5_2_161_0}. Montgomery
\cite{0951-7715-11-2-011} discovered a family of periodic solutions,
One of which is presented in fig.\,~\ref{fig:braids}.

Newton could, in principle, have realized that stable periodic
solutions existed. Such solutions, called braids, are found by
laboriously searching parameter space using action minimizers under
topological constraints
\cite{2004NYASA1017..422V,2015IJBC...2550169Y}. Braids are stable,
regular, and once found, they can be integrated relatively easily
\cite{2000MNRAS.318L..61H}. Due to a lack of computer resources,
Newton would not have been able to calculate these solutions, even
during his two years of social distancing.

Non-periodic democratic solutions, as depicted in
fig.\,~\ref{fig:braids}, on the other hand, are impossible to solve by
hand \cite{1983ApJ...268..319H}, and even with modern digital
computers they can only be approximated \cite{2020MNRAS.493.3932B}.

\section*{The paradox of infinitesimal granularity}

The intrinsic chaotic nature of the democratic N-body problem causes
small errors introduced by computer round-off and integration time
discretizations to grow exponentially \cite{2016ttp..book.....V}.
Still, solutions can be acquired by systematically reducing the time
step and increasing the computer's mantissa until the solution
converges. Finding such a solution is an elaborate iterative process
in which the time step is reduced and the mantissa extended until the
result becomes independent of these.

For this purpose, we developed {\tt Brutus}, a direct $N$-body solver
for self-gravitating systems of $N$ point-masses to arbitrary
precision \cite{PORTEGIESZWART2018160}. In {\tt Brutus} round-off is
controlled by specifying the word-length $L_w$ in which floating
point arithmetic is performed using the MPFR arbitrary-precision
library \cite{2014arXiv1402.6713P}. To assure that the method of
numerical integration does not introduce a bias in our simulations we
adopted the Gragg-Bulirsch-Stoer algorithm
\cite{springerlink:10.1007/BF01386092}, which combines the modified
midpoint method \cite{Gragg1965} for solving Newton's ordinary
differential equation \cite{Newton:1687} with Richardson extrapolation
\cite{Richardson1911} to improve the rate of convergence. We can now
control the algorithmic error by changing an energy tolerance
parameter $\epsilon$ and the length of the mantissa $L_w$: Both
parameters can be varied simultaneously as in
\cite{2018CNSNS..61..160P}
\begin{equation}
L_w = 4 |\log_{10}(\epsilon)| + 32 \, \textrm{bits}.
\end{equation}

By repeating calculations with identical realizations of the initial
conditions while systematically changing $\epsilon$ and $L_w$ we
can achieve a solution that becomes independent of $\epsilon$ and
$L_w$ to $p$ decimal places.

Integrating converged solutions scales $\propto N^2$, but requires
many iterations, which progressively become more expensive. Acquiring
a converged solutions to an $N$-body problem is easily $10^4$ times
more expensive than regular integration techniques using IEEE~754
double precision arithmetic \cite{2022A&A...659A..86P}. Any
non-converged solution to a self-gravitating system under Newton's
equations of motion should be subject to considerable concern
\cite{1964ApJ...140..250M}, although statistically, the phase-space
coverage of the solutions may be indistinguishable from the true
outcome space \cite{2015ComAC...2....2B}.

\begin{figure}
\centering
\includegraphics[width=1.0\textwidth]{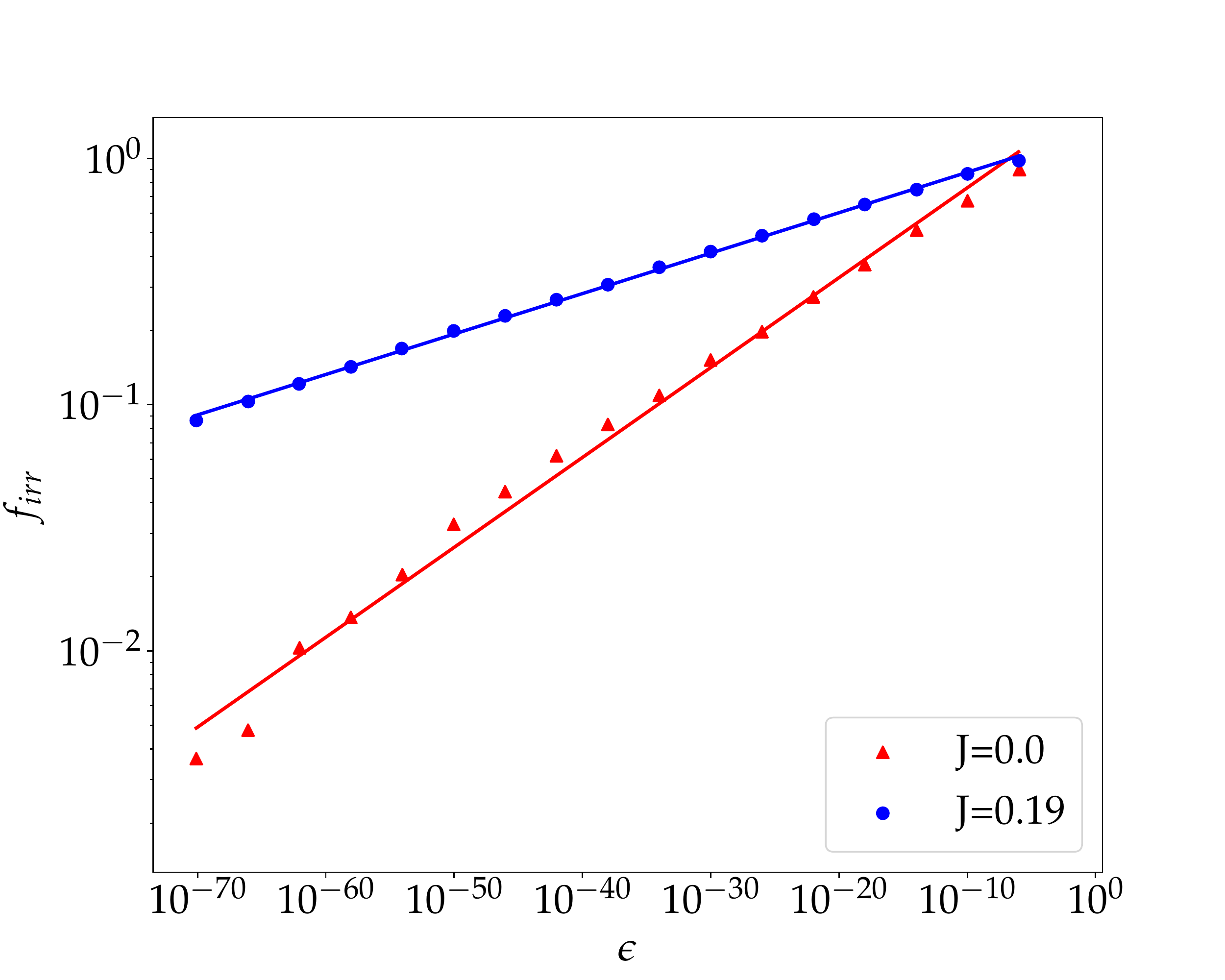}
\caption{The fraction of irreversible systems of converged solutions
for zero-angular momentum-triples ($J=0.0$ in red) and triples with
the maximum amount of angular momentim giving virial initial
conditions ($J=0.19$ in blue). Both curves can be fitted by a power
law: $\log_{10}\,f_{\rm{irr}} \simeq 0.036 \log_{10}\,\epsilon +
0.25$ for $J=0$, and $\log_{10}\,f_{\rm{irr}} \simeq 0.016
\log_{10}\,\epsilon + 0.11$. }
\label{fig:time_reversibility_fraction}
\end{figure}

The measure of chaos, often quantified with the largest positive
Lyapunov exponent expressed in a time scale, $t_{\rm Ly}$, cannot be
determined without performing extensive numerical analyses. We do not
know how chaotic $N$-body systems are, but estimate values for
$t_{\rm Ly}$ by running the same initial realization twice until a
converged solution is achieved: In the second calculation, a small
relative displacement of a random particle by $10^{-10}$ is introduced
in one of the principal Cartesian coordinates. We calculate the
systems until a single star escapes. We then turn time backward in
the computer, and calculate the final conditions back to their initial
realization, recovering the $10^{-10}$ initial displacement of the
perturbed solution with respect to the unperturbed solution
\cite{2018CNSNS..61..160P}. When performing this operation for a
large number of randomly generated triples, we can calculate how many
of those converged solutions were successfully run backward as a
function of $\epsilon$. In
figure\,\ref{fig:time_reversibility_fraction} we present this fraction
of irregular (or irreturnable) solutions $f_{\rm irr}$ as a function
of $\epsilon$.

We were surprised to find that triples with finite angular momentum
turn out to be more chaotic than their zero-angular momentum
counterparts. The underlying reason hides in the triple's lifetime;
high-angular momentum triples have a lower rate of close encounters,
and those encounters are typically at a larger mutual distance. As a
consequence, low angular momentum systems tend to have a higher
probability of violent, stellar ejecting encounters causing these
systems to live shorter on average compared to their high
angular-momentum equivalents. The shorter lifetime of low-angular
momentum systems cause them to have less time for chaos to propagate,
even at the same Lyapunov time scale.

Another remarkable aspect of
figure\,\ref{fig:time_reversibility_fraction} is that about $\apgt
10$\% of virialized triples are so chaotic that these trajectories are
affected by perturbations on a relative scale of $10^{-70}$
\cite{2020MNRAS.493.3932B}. When considering the size of the
universe, a perturbation of $1.3 \cdot 10^{-34}$\,$\angstrom$, or
$\sim 8.1 \cdot 10^{-10}$\,lp would then still drive exponential
divergence to a level of making the system unpredictable at any scale.

\section*{Global chaos due to Newton's equations of motion}

Newton operates on all scales and independently of the intrinsic
nature of the objects, either being black holes or the balls of a
juggler. The interference of other physical processes is irrelevant
so long as the dynamics is driven in part by Newton's equations of
motion. The fundamental irreversibility of time in Newton's equations
of motion propagates to all scales and affects every object
\cite{1964ApJ...140..250M}. Time irreversibility on macroscopic
scales then is a consequence of the chaotic nature of Newton's
equation of motion, as the spontaneous collapse of the wave function
\cite{PhysRevD.34.470} introduces stochastic processes on a
microscopic scale that propagate to macroscopic scales.

\begin{figure}
\centering
\includegraphics[width=1.0\columnwidth]{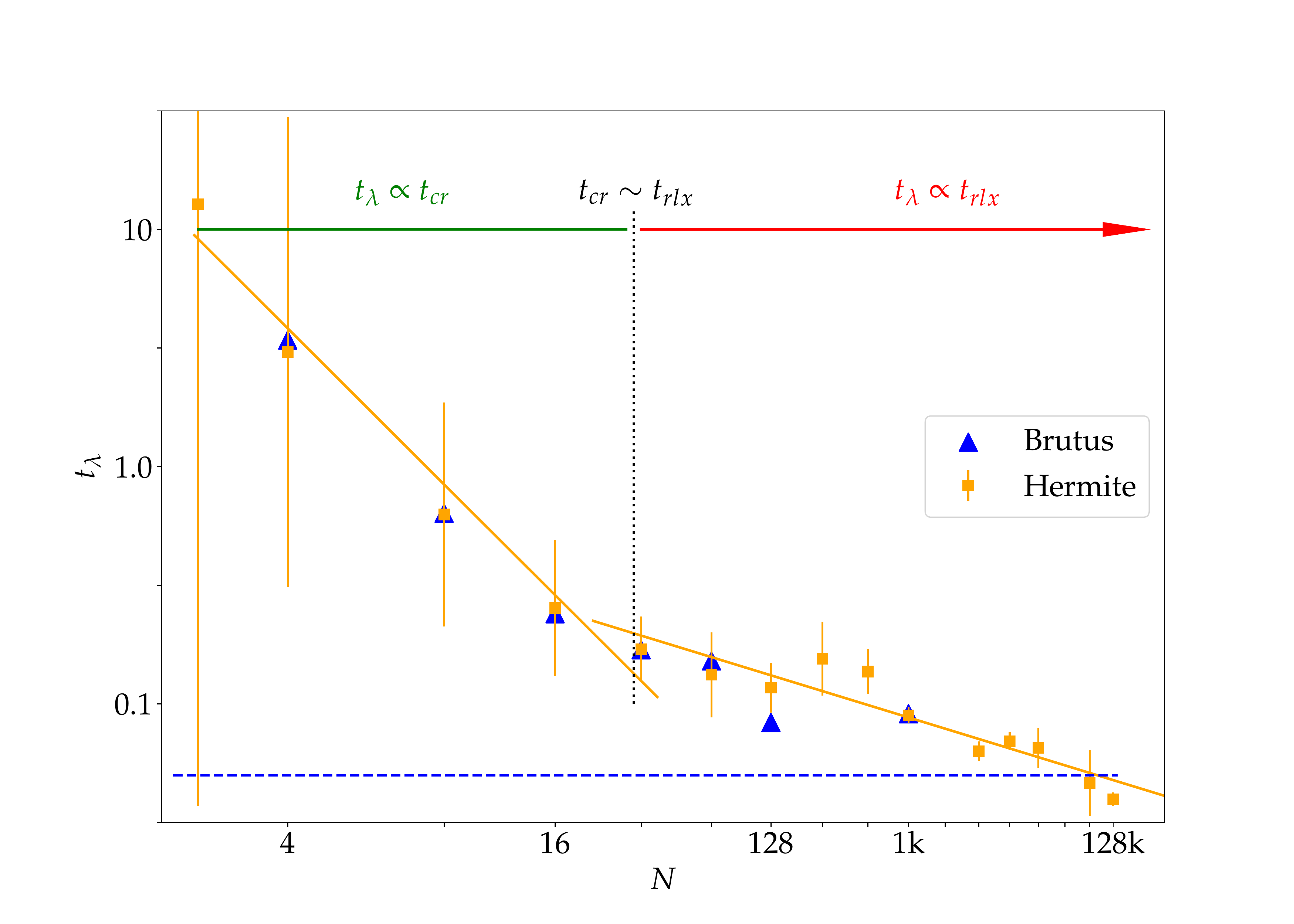}
\caption{Estimate of the Lyapunov timescale as a function of the
number of particles. Here the horizontal axis is not linear, but in
$\ln(\ln(N))$ to illustrate the scaling proposed in
\cite{1993ApJ...415..715G}. The different symbols and colors
represent different calculations (see legend). The vertical bars,
plotted for Newton's Hermite only, show the root-mean-square of the
dispersion in the series of solutions. The error bars in the results
obtained with Brutus are statistically indistinguishable from the
presented bars. The blue horizontal dotted line accentuates the
boundary where large $N$ systems become on average as chaotic as the
most chaotic 3-body systems in our adopted initial conditions.
}
\label{fig:Newton's_Lyapunov_timescale}
\end{figure}

After exploring the three-body problem, we continue with virialized
equal-mass large $N$-systems in a homogeneous unit cube (we adopted
dimension-less coordinates in units with $G=1$
\cite{1971Ap&SS..13..284H,1986LNP...267..233H}). We explore systems
with up to $N=131072$ ($\equiv 128$\,k). The results are presented in
fig\,\ref{fig:Newton's_Lyapunov_timescale}, where we show the Lyapunov
time scale for of ensembles of multi-body systems. These calculations
are continued for about 3 or 4 dynamical times, but the measurements
are carried out on the sub-sample of the data, from $t=1/\pi$ until
growth of the phase-space distance between two converged solutions of
a perturbed and a non-perturbed initial realization exceeds 0.1. In
some cases, the exponential growth of the phase-space distance
eventually flattens to a $\log_{10}(\delta) \propto t$ scaling
relation (consistent with the expectation for a relaxation-driven
process, see \cite{2002JSP...109.1017H}). We acquired converged
solutions for $N=4$, 8, 16, 32, 64, 128, and $N=1024$, and
reprehensive \cite{2018CNSNS..61..160P} solutions for $N$ up to 128k
particles using a regular fourth-order Hermite integration algorithm
using IEEE 754 double precision arithmetic.

The Lyapunov time-scale $\propto \ln(\ln(N))$
\cite{1993ApJ...415..715G} with a discontinuity near $N=32$ due to
the transition from crossing-time dominated to relaxation time
dominated dynamics. The distribution width of Lyapunov timescales for
larger $N$ is narrower than for small $N$. For up to $N\sim 128$\,k
the majority of systems tend to be more regular than for $N=3$. For
larger $N$, however, all systems tend to be even more chaotic than the
most chaotic three-body systems.

We argue that if wave functions collapse spontaneously, as is the case
in Ghirardi-Rimini-Weber theory \cite{PhysRevD.34.470}, microscopic
stochastic effects propagate non-linearly to macroscopic scales
through Newtonian dynamics. The motion of a considerable fraction of
triples of self-gravitating bodies, and possibly the vast majority of
systems with $\apgt 128$\,k are affected by the spontaneous collapse
of the wave function. As a consequence, the evolution of
self-gravitating multi-body system is eventually stochastic, due to
the proliferation of chaotic processes that propagate from microscopic
scales to macroscopic scales. This connection between the microscopic
quantum scale and macroscopic scales cause the dynamics to be
fundamentally time irreversible; Newton's equations of motion are
incalculable (they cannot be calculated precisely in a deterministic
nature) and inconsequential (there is no point in achieving such a
result).

Changes in the macroscopic distribution of mass in the Galaxy are
driven by chaos on microscopic scales.  Due to the spontaneous
collapse of the wave function, perturbations at Planck length scales
are magnified exponentially by the chaotic dynamics of planetary
systems, stellar multiples, star clusters and giant molecular clouds.
This growth cannot continue indefinitely, and converges once the
perturbation has approximately reached the size of the Galaxy
\cite{2000chun.proc..229V}.  The divergence in orbital energy then
starts exponentially but eventually transforms into a much slower
random walk, consistent with the global relaxation process
\cite{2002JSP...109.1017H,2002ApJ...580..606H}.  Hence, one may think
of a Galactic butterfly effect, where fluctuations lead to
unpredictable local events, but which only gradually affect the global
Galactic climate.

\section*{Acknowledgments}

It is a pleasure to thank David Albert and Douglas Heggie for
discussions.


\begin{thebibliography}{10}
\providecommand{\url}[1]{\texttt{#1}}
\providecommand{\urlprefix}{URL }
\expandafter\ifx\csname urlstyle\endcsname\relax
  \providecommand{\doi}[1]{doi:\discretionary{}{}{}#1}\else
  \providecommand{\doi}{doi:\discretionary{}{}{}\begingroup
  \urlstyle{rm}\Url}\fi
\providecommand{\eprint}[2][]{\url{#2}}

\bibitem{1891BuAsI...8...12P}
H.~{Poincar{\'e}},
\newblock \emph{{M{\'e}moires et observations. Sur le probl{\`e}me des trois
  corps}},
\newblock Bulletin Astronomique, Serie I \textbf{8}, 12 (1891).

\bibitem{zbMATH02627145}
K.~F. {Sundman},
\newblock \emph{{M\'emoire sur le probl\`eme des trois corps}},
\newblock {Acta Math.} \textbf{36}, 105 (1912).

\bibitem{zbMATH02639473}
K.~F. {Sundman},
\newblock \emph{{Recherches sur le probl\`eme des trois corps. Nouvelles
  recherches su r le probl\`eme des trois corps}},
\newblock {Acta Soc. Sc. Fennicae 34, No. 6, 1-43 (1907); 35, No. 9, 1-27
  (1909).} (1909).

\bibitem{PHSC_2001__5_2_161_0}
M.~Henkel,
\newblock \emph{Sur la solution de sundman du probl\`eme des trois corps},
\newblock Philosophia Scientiae \textbf{5}(2), 161 (2001).

\bibitem{0951-7715-11-2-011}
R.~Montgomery,
\newblock \emph{The n -body problem, the braid group, and action-minimizing
  periodic solutions},
\newblock Nonlinearity \textbf{11}(2), 363 (1998).

\bibitem{2004NYASA1017..422V}
R.~J. {Vanderbei},
\newblock \emph{{New Orbits for the n-Body Problem}},
\newblock Annals of the New York Academy of Sciences \textbf{1017}(1), 422
  (2004),
\newblock \doi{10.1196/annals.1311.024},
\newblock \eprint{astro-ph/0303153}.

\bibitem{2015IJBC...2550169Y}
D.~{Yan}, R.~{Liu}, X.~{Hu}, W.~{Mao} and T.~{Ouyang},
\newblock \emph{{New Phenomena in the Spatial Isosceles Three-Body Problem with
  Unequal Masses}},
\newblock International Journal of Bifurcation and Chaos \textbf{25}(12),
  1550169-313 (2015),
\newblock \doi{10.1142/S0218127415501692}.

\bibitem{2000MNRAS.318L..61H}
D.~C. {Heggie},
\newblock \emph{{A new outcome of binary-binary scattering}},
\newblock \mnras \textbf{318}(4), L61 (2000),
\newblock \doi{10.1046/j.1365-8711.2000.04027.x},
\newblock \eprint{astro-ph/9604016}.

\bibitem{2018PASJ...70...64L}
X.~{Li}, Y.~{Jing} and S.~{Liao},
\newblock \emph{{Over a thousand new periodic orbits of a planar three-body
  system with unequal masses}},
\newblock \pasj \textbf{70}(4), 64 (2018),
\newblock \doi{10.1093/pasj/psy057},
\newblock \eprint{1709.04775}.

\bibitem{1983ApJ...268..319H}
P.~{Hut} and J.~N. {Bahcall},
\newblock \emph{{Binary-single star scattering. I - Numerical experiments for
  equal masses}},
\newblock \apj \textbf{268}, 319 (1983),
\newblock \doi{10.1086/160956}.

\bibitem{2020MNRAS.493.3932B}
T.~C.~N. {Boekholt}, S.~F. {Portegies Zwart} and M.~{Valtonen},
\newblock \emph{{Gargantuan chaotic gravitational three-body systems and their
  irreversibility to the Planck length}},
\newblock \mnras \textbf{493}(3), 3932 (2020),
\newblock \doi{10.1093/mnras/staa452},
\newblock \eprint{2002.04029}.

\bibitem{2016ttp..book.....V}
M.~{Valtonen}, J.~{Anosova}, K.~{Kholshevnikov}, A.~{Myll{\"a}ri}, V.~{Orlov}
  and K.~{Tanikawa},
\newblock \emph{{The Three-body Problem from Pythagoras to Hawking}},
\newblock \doi{10.1007/978-3-319-22726-9} (2016).

\bibitem{PORTEGIESZWART2018160}
S.~F. {Portegies Zwart} and T.~C. Boekholt,
\newblock \emph{Numerical verification of the microscopic time reversibility of
  newton’s equations of motion: Fighting exponential divergence},
\newblock Communications in Nonlinear Science and Numerical Simulation
  \textbf{61}, 160  (2018),
\newblock \doi{https://doi.org/10.1016/j.cnsns.2018.02.002}.

\bibitem{2014arXiv1402.6713P}
S.~{Portegies Zwart} and T.~{Boekholt},
\newblock \emph{{On the minimal accuracy required for simulating
  self-gravitating systems by means of direct N-body methods}},
\newblock ApJL in press, ArXiv e-prints 1402.6713  (2014),
\newblock \eprint{1402.6713}.

\bibitem{springerlink:10.1007/BF01386092}
R.~Bulirsch and J.~Stoer,
\newblock \emph{Fehlerabschätzungen und extrapolation mit rationalen
  funktionen bei verfahren vom richardson-typus},
\newblock Numerische Mathematik \textbf{6}, 413 (1964),
\newblock 10.1007/BF01386092.

\bibitem{Gragg1965}
W.~Gragg,
\newblock \emph{On extrapolation algorithms for ordinary initial value
  problems},
\newblock Journal of the Society for Industrial and Applied Mathematics
  \textbf{2}(3), 384 (1965).

\bibitem{Newton:1687}
I.~Newton,
\newblock \emph{Philosophiae Naturalis Principia Mathematica}, vol.~1 (1687).

\bibitem{Richardson1911}
L.~{Richardson},
\newblock \emph{{The Approximate Arithmetical Solution by Finite Differences of
  Physical Problems Involving Differential Equations, with an Application to
  the Stresses in a Masonry Dam}},
\newblock Phil. Trans. R. Soc. Lond. A \textbf{210}(459-470), 307 (1911).

\bibitem{2018CNSNS..61..160P}
S.~F. {Portegies Zwart} and T.~C.~N. {Boekholt},
\newblock \emph{{Numerical verification of the microscopic time reversibility
  of Newton's equations of motion: Fighting exponential divergence}},
\newblock Communications in Nonlinear Science and Numerical Simulations
  \textbf{61}, 160 (2018),
\newblock \doi{10.1016/j.cnsns.2018.02.002},
\newblock \eprint{1802.00970}.

\bibitem{2022A&A...659A..86P}
S.~F. {Portegies Zwart}, T.~C.~N. {Boekholt}, E.~H. {Por}, A.~S. {Hamers} and
  S.~L.~W. {McMillan},
\newblock \emph{{Chaos in self-gravitating many-body systems. Lyapunov time
  dependence of N and the influence of general relativity}},
\newblock \aap \textbf{659}, A86 (2022),
\newblock \doi{10.1051/0004-6361/202141789},
\newblock \eprint{2109.11012}.

\bibitem{1964ApJ...140..250M}
R.~H. {Miller},
\newblock \emph{{Irreversibility in Small Stellar Dynamical Systems.}},
\newblock \apj \textbf{140}, 250 (1964),
\newblock \doi{10.1086/147911}.

\bibitem{2015ComAC...2....2B}
T.~{Boekholt} and S.~{Portegies Zwart},
\newblock \emph{{On the reliability of N-body simulations}},
\newblock Computational Astrophysics and Cosmology \textbf{2}, 2 (2015),
\newblock \doi{10.1186/s40668-014-0005-3},
\newblock \eprint{1411.6671}.

\bibitem{PhysRevD.34.470}
G.~C. Ghirardi, A.~Rimini and T.~Weber,
\newblock \emph{Unified dynamics for microscopic and macroscopic systems},
\newblock Phys. Rev. D \textbf{34}, 470 (1986),
\newblock \doi{10.1103/PhysRevD.34.470}.

\bibitem{1971Ap&SS..13..284H}
M.~{H{\'e}non},
\newblock \emph{{Monte Carlo Models of Star Clusters (Part of the Proceedings
  of the IAU Colloquium No. 10, held in Cambridge, England, August 12-15,
  1970.)}},
\newblock \apss \textbf{13}, 284 (1971).

\bibitem{1986LNP...267..233H}
D.~C. {Heggie} and R.~D. {Mathieu},
\newblock \emph{{Standardised Units and Time Scales}},
\newblock In P.~{Hut} and S.~L.~W. {McMillan}, eds., \emph{The Use of
  Supercomputers in Stellar Dynamics}, vol. 267 of \emph{Lecture Notes in
  Physics, Berlin Springer Verlag}, p. 233,
\newblock \doi{10.1007/BFb0116419} (1986).

\bibitem{2002JSP...109.1017H}
P.~{Hut} and D.~C. {Heggie},
\newblock \emph{{Orbital Divergence and Relaxation in the Gravitational N-Body
  Problem}},
\newblock Journal of Statistical Physics \textbf{109}(5-6), 1017 (2002),
\newblock \doi{10.1023/A:1020472526203},
\newblock \eprint{astro-ph/0111015}.

\bibitem{1993ApJ...415..715G}
J.~{Goodman}, D.~C. {Heggie} and P.~{Hut},
\newblock \emph{{On the Exponential Instability of N-Body Systems}},
\newblock \apj \textbf{415}, 715 (1993),
\newblock \doi{10.1086/173196}.

\bibitem{2000chun.proc..229V}
M.~{Valluri} and D.~{Merritt},
\newblock \emph{{Orbital Instability and Relaxation in Stellar Systems}},
\newblock In V.~G. {Gurzadyan} and R.~{Ruffini}, eds., \emph{The Chaotic
  Universe}, pp. 229--246,
\newblock \doi{10.1142/9789812793621\_0014} (2000), \eprint{astro-ph/9909403}.

\bibitem{2002ApJ...580..606H}
M.~{Hemsendorf} and D.~{Merritt},
\newblock \emph{{Instability of the Gravitational N-Body Problem in the Large-N
  Limit}},
\newblock \apj \textbf{580}(1), 606 (2002),
\newblock \doi{10.1086/343027},
\newblock \eprint{astro-ph/0205538}.

\end{thebibliography}
\end{document}